\documentclass{article}

\usepackage{microtype}
\usepackage{multirow}
\usepackage{enumitem}
\usepackage{booktabs}
\usepackage{threeparttable}
\usepackage{graphicx}
\usepackage{subfigure}
\usepackage{listings}
\usepackage{xcolor}
\usepackage{changepage}
\usepackage{float}

\usepackage{longtable}

\lstdefinestyle{feedbackprompt}{
    basicstyle=\ttfamily\small,
    breaklines=true,
    columns=fullflexible,
    keepspaces=true,
    frame=single,
}
\usepackage[utf8]{inputenc}
\usepackage{booktabs} 

\usepackage{hyperref}

\usepackage[accepted]{icml2026}

\usepackage{amsmath}
\usepackage{amssymb}
\usepackage{mathtools}
\usepackage{amsthm}

\usepackage[capitalize,noabbrev]{cleveref}

\theoremstyle{plain}

\theoremstyle{definition}

\theoremstyle{remark}

\icmltitlerunning{Stable Personas}

\begin{document}

\twocolumn[
\icmltitle{Stable Personas: \\ Dual-Assessment of Temporal Stability in LLM-Based Human Simulation}

\begin{icmlauthorlist}
\icmlauthor{Jana Gonnermann-Müller}{1,2}
\icmlauthor{Jennifer Haase}{2,3}
\icmlauthor{Nicolas Leins}{1}
\icmlauthor{Thomas Kosch}{3,2}
\icmlauthor{Sebastian Pokutta}{1,4}
\end{icmlauthorlist}

\icmlaffiliation{1}{Zuse Institute Berlin, Berlin, Germany}
\icmlaffiliation{2}{Weizenbaum Institute, Berlin, Germany}
\icmlaffiliation{3}{Humboldt-University Berlin, Berlin, Germany}
\icmlaffiliation{4}{Technical University Berlin, Berlin, Germany}

\icmlcorrespondingauthor{Jana Gonnermann-Müller}{gonnermann-mueller@zib.de}

\icmlkeywords{LLM, Stability, Human Simulation, Social Science, Variance Secomposition}

\vskip 0.3in
]

\printAffiliationsAndNotice{}  

\begin{abstract}
Large Language Models (LLMs) acting as artificial agents offer the potential for scalable behavioral research, yet their validity depends on whether LLMs can maintain stable personas across extended conversations. We address this point using a dual-assessment framework measuring both self-reported characteristics and observer-rated persona expression. Across two experiments testing four persona conditions (default, high, moderate, and low ADHD presentations), seven LLMs, and three semantically equivalent persona prompts, we examine between-conversation stability ($3,473$ conversations) and within-conversation stability ($1,370$ conversations × $18$ turns). 
Self-reports remain highly stable both between and within conversations. However, observer ratings reveal a tendency for persona expressions to decline during extended conversations. These findings suggest that persona-instructed LLMs produce stable, persona-aligned self-reports, an important prerequisite for behavioral research, while identifying this regression tendency as a boundary condition for multi-agent social simulation.
\end{abstract}

\section{Introduction}
\label{Introduction}
Large language models (LLMs) configured as artificial agents enable scalable research through simulations of diverse human populations \cite{hu_simbench_2026, murthy_one_2025, binz_foundation_2025, aher_using_2023}, with applications spanning educational tutoring \cite{mannekoteCanLLMsReliably2025, li_can_2025}, prototype testing \cite{hamalainen_evaluating_2023}, and therapeutic dialogue \cite{hu_theramind_2025}. Recent advances toward agentic AI systems capable of reasoning, planning, and coordination \cite{anthis_position_2025, haase_beyond_2025} have accelerated progress in multi-agent simulations for social research. However, these applications rest on an untested assumption \emph{that LLMs maintain stable personas across extended conversations}. This assumption has direct consequences for simulation validity. Social phenomena such as negotiation, trust formation, and group dynamics emerge through sustained path-dependent interactions. These interactions are shaped by memory and mutual adaptation. An agent whose personality drifts mid-conversation or shifts unpredictably across sessions cannot meaningfully participate in such processes. If constituent agents lack temporal stability, emergent patterns become artifacts of model inconsistency rather than genuine social phenomena. Establishing persona stability is, therefore, a necessary prerequisite for meaningful multi-agent social simulation \cite{rozen_llms_2025,abdulhai_consistently_2025}.

The evidence remains mixed. Studies demonstrate human-like sensitivity to incentive structures \cite{nguyen_navigating_2025}, trust dynamics in repeated games \cite{xie_can_2024}, and internally stable personality questionnaire responses \cite{jiang_personallm_2024}. Recent work on unified cognitive models suggests that LLMs may capture behavioral regularities across diverse psychological domains \cite{binz_foundation_2025}. Yet, models also exhibit sycophancy toward socially desirable responses \cite{salminen_deus_2024,sharma_towards_2024}, systematically overestimating human rationality \cite{liu_large_2025}, and drifting toward an ``average persona'' over time \cite{taillandier_integrating_2025}. 
These limitations may reflect architectural constraints, as LLMs exhibit working memory capacity limits analogous to those of humans \cite{gong_working_2024}, suggesting that persona stability faces similar bottlenecks. 

Current evaluation paradigms fail to capture stability aspects. Most research treats LLM output as a single-trial result, implicitly assuming reproducibility while neglecting to test this assumption or examine design decisions such as model choice and prompting strategy \cite{cummins_threat_2025}. 
Moreover, reliance on single-source assessments cannot detect dissociations between internal persona representations and persona expression, which is a critical distinction when simulated therapists, tutors, or research participants interact with humans who form trust based on the perceived consistency of behavior. 

We address these gaps through three methodological advances. 
First, we introduce a dual-assessment framework adapted from clinical psychology, jointly measuring self-reported characteristics and observer-rated persona expressions to detect stability failures that are invisible to single-source evaluation \cite{olino_psychometric_2015}. 
Second, we operationalize personas across the full spectrum using clinically-grounded ADHD profiles, which we use as a test case that deviates from normative training distributions and exposes sycophancy and tendencies toward average responses. We define persona as a specified configuration of characteristics, behavior, and goals that an LLM is instructed to embody.  
Third, we quantify both between-conversation and within-conversation stability across seven models and three equivalent prompt designs, enabling systematic variance decomposition that distinguishes persona effects from model- and prompt-induced variability, addressing the following research question: 

\begin{adjustwidth}{1.5em}{}
    \textit{How stably do LLMs maintain assigned personas across independent conversations and throughout extended conversations?}
\end{adjustwidth}

Two experiments assess two dimensions of temporal stability: 
Experiment~I tests between-conversation stability ($N=3,473$, 50 independent runs per condition); 
Experiment~II tests within-conversation stability ($N=1,370$ conversations with 18-turn conversations). 

We control for (1)~model choice, testing seven LLMs spanning proprietary and open-source systems; (2)~prompt design, varying three semantically equivalent formats.

\textbf{Our Contributions:}
\begin{enumerate}[itemsep=1pt, parsep=0pt, topsep=1pt, partopsep=0pt, leftmargin=*]
    \item \textbf{Stable self-reports, bounded persona expression.} We demonstrate that LLMs produce stable, persona-aligned self-reports. Persona observer ratings of high and moderate intensities decline over extended conversations, indicating a boundary condition for multi-agent social simulation. 
    
    \item \textbf{Large-scale stability analysis.} We provide systematic evidence on persona stability across seven models, three prompts, and four persona intensities, assessing both between-conversation and within-conversation stability.

   \item \textbf{Dual-assessment framework.} We adapt a multi-informant methodology to evaluate LLMs by jointly measuring self-reported characteristics and observer-rated persona expressions, which are invisible to single-source approaches.
\end{enumerate}

\section{Methodology}
\label{sec:methodology}

\subsection{Experiment Procedure}

\begin{figure*}[t]
    \centering
    \includegraphics[width=0.8\textwidth]{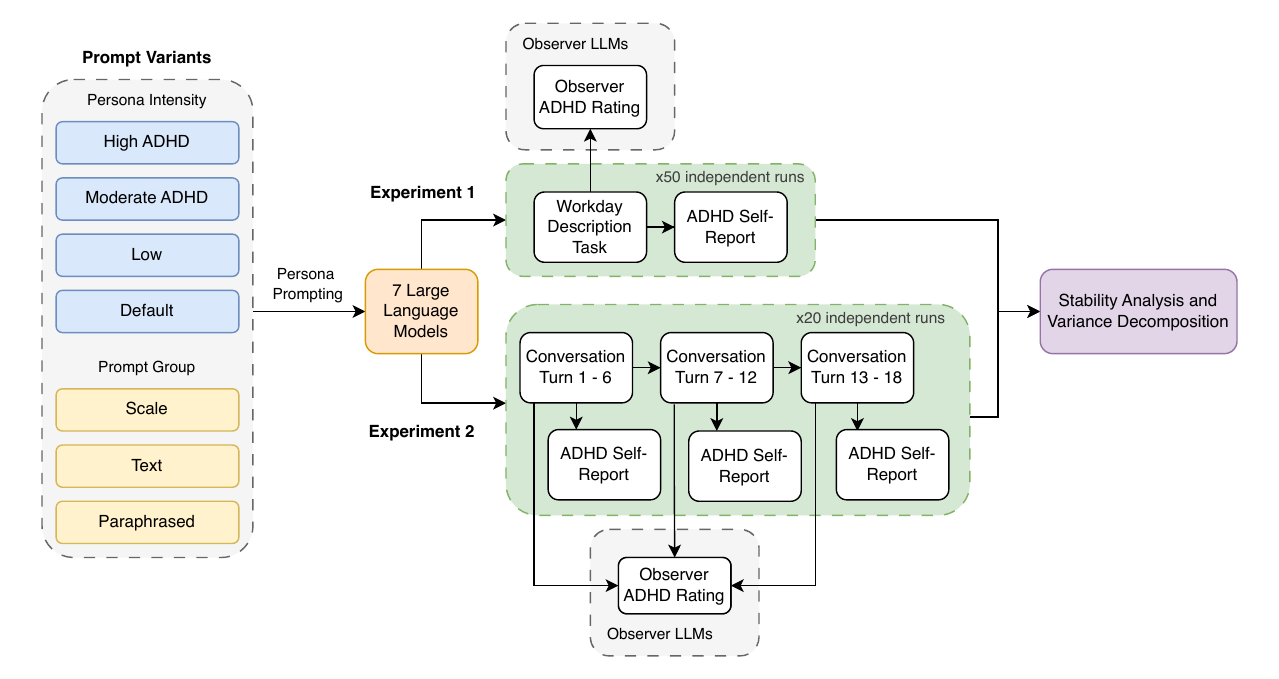}
    \caption{Procedure for Experiments I and II (combined presentation).}
    \label{fig:experiment-pipeline}
    \vspace{-1em}
\end{figure*}

We operationalize stability through two complementary experiments (Figure \ref{fig:experiment-pipeline}). \textit{Experiment~I} tests between-conversation stability across 50 independent instantiations per condition. Each model generates a first-person workday narrative and completes an ADHD self-report scale, while three LLM raters evaluate the narratives using an observer-report scale without access to persona instructions (see the task in Appendix~\ref{app:experiment_task}). \textit{Experiment~II} tests within-conversation stability. Specifically, the target LLM engages in an 18-turn neutral conversation, with assessments at three time points (turns 6, 12, and 18). The neutral conversation partner is instantiated to listen without judgment and to ask simple questions that help the speaker continue talking without sharing opinions or trying to influence what they say (see the full instruction in Appendix~\ref{app:conversationpartner}). The models complete self-reports, while LLM evaluators rate the accumulated turns. 

\subsection{Experiment Factors}

\paragraph{Persona intensity.}
We operationalize behavior using diagnostic criteria for Attention Deficit/Hyperactivity Disorder (ADHD) derived from the DSM-5 and ICD-10 \cite{world_health_organisation_who_international_2025,american_psychiatric_association_-_apa_diagnostisches_2025}, the international diagnostic and classification catalogs for psychological assessment. Based on the criteria, we employ four conditions: \textit{high}, \textit{moderate}, and \textit{low} expressions of ADHD characteristics, using the \textit{default} as a control with no persona description. See all persona prompts in Appendix Table~\ref{tab:persona_prompts}. 

\paragraph{Model choice.}
We utilized seven models spanning proprietary and open-source LLMs: Claude Sonnet 4.5, DeepSeek V3.2, GPT 5.1, GPT OSS 120B, Gemini 3 Pro, Grok 4.1, and Llama 3.3 70B. The selection captures a broad cross-section of the field, including large-scale proprietary systems from US and Chinese firms, along with prominent open-source alternatives. Proprietary models were queried through official provider APIs; open-weight models ran locally via Ollama. We used provider-default decoding settings (temperature, top-p, top-k) to reflect typical user-facing behavior. Consequently, model effects conflate architecture and default decoding/alignment choices (for an overview, see Table~\ref{tab:LLMs} in the Appendix).

\paragraph{Prompt design.}
We use three semantically equivalent prompts conveying identical content: a \textit{text-based prompt} specifying characteristics, role, and goal, with persona intensity expressed through frequency adverbs; a \textit{scale-based prompt} encoding the same attributes using 7-point Likert ratings \cite{jebb_review_2021}; and a control condition with \textit{rephrased/reworded} semantically equivalent phrasing and altered information order (see Appendix Table~\ref{tab:persona_prompts}).

\subsection{Measurement}

We use the Conners' Adult ADHD Rating Scales (CAARS) \cite{conners_conners_nodate}, a validated instrument with parallel self-report and observer-report forms. We extract the 12-item ADHD Index (range: $0-36$) as our primary measure.

Three independent LLM raters (Claude Sonnet 4.5, GPT-5.1, Gemini 3 Pro) independently rated each narrative. We selected three instruction-following LLMs from different providers to reduce reliance on a single model family. 
Observer-report scores were aggregated across three independent LLM evaluators. Inter-rater reliability was high, with an Intraclass Correlation Coefficient ICC(2,1) of $.90$ ($95\%$ CI $[.70–.95]$) across $N=3,473$ conversations (Exp~I), permitting aggregation into mean observer scores. Similar inter-rater reliability was observed within conversations (Exp~II) with ICC(2,1) = $.83$ ($95\%$ CI $[.51-.92]$) at turn 6, ICC(2,1) = $.75$ ($95\%$ CI $[.38-.87]$) at turn 12, and ICC(2,1) = $.69$ ($95\%$ CI $[.32-.84]$) at turn 18.

While high inter-rater reliability among LLM judges demonstrates rater agreement, it does not establish construct validity, as consistent ratings may stem from shared model biases rather than accurate ADHD assessment. Therefore, to verify that LLM ratings capture meaningful persona expression, we compared LLM observer ratings to human expert judgments using a sample of 20 narratives (6 high ADHD, 6 moderate ADHD, 6 low, 2 default) drawn from two target models (Claude Sonnet 4.5, Gemini 3 Pro) across text-based and scale-based prompting formats. Five M.Sc.-level psychologists independently rated all narratives using the identical observer scale of the CAARS, blind to the persona condition. Human inter-rater reliability showed excellent consistency, with a two-way random-effects model, absolute agreement, single measure ICC(2,1) = $.92$ ($95\%$ CI $[.85-.96]$). LLM inter-rater reliability was lower than human reliability but showed good overall agreement ICC(2,1) = $.87$, ($95\%$ CI $[.55-.95]$). Human-LLM convergent validity, calculated using ICC(2,1) between aggregated mean ratings, was excellent ICC(2,1) = $.95$, ($95\%$ CI $[.57-.99]$). 

\subsection{Data Analysis}
Data collection occurred in December 2025. We compute independent ADHD intensity ratings from both self-report and observer-report forms, with scores ranging from $0-36$; higher values indicate greater ADHD symptom intensity. The 12-item ADHD Index serves as our primary outcome measure across all analyses.

Experiment~I comprised $N = 3,473$ single-turn conversations (7 models $\times$ 4 personas $\times$ 3 prompts $\times$ 50 runs + 7 models $\times$ 50 runs with default configuration) with $N=10,419$ observer assessments and $N=3,473$ self-reports. Experiment~II comprised $N = 1,370$ multi-turn conversations (7 models $\times$ 4 personas $\times$ 3 prompts + 7 models $\times$ 20 runs with the default configuration). Each conversation extended to 18 turns with assessments at three checkpoints, yielding $N = 12,201$ observer assessments (turn 6: $n = 4,062$; turn 12: $n = 4,047$; turn 18: $n = 4,092$) and $N = 4,054$ self-report measurements (turn 6: $n = 1,343$; turn 12: $n = 1,341$; turn 18: $n = 1,370$). The final sample sizes are slightly lower than the theoretical maximums due to minor data attrition. A small number of runs (0.77\% for Experiment~I, 2.14\% for Experiment~II) could not be generated completely due to API connectivity interruptions or malformed JSON outputs of questionnaire items that prevented automated parsing. For a full list, see Table~\ref{tab:cell_counts} in the Appendix.

We quantify temporal stability using the standard deviation ($SD$) of ADHD Index scores within each experimental condition, where higher variability indicates lower stability. We conducted separate analyses for self-report and observer-report measures. To partition sources of variance, we fit linear mixed-effects models with the ADHD Index score $Y_{mpq}$ (persona $p$, model $m$, prompt $q$) as the dependent variable. 
For Experiment~I: $\mu$ is the grand mean, $\alpha_p$ is the fixed effect for persona, $\beta_m \sim N(0, \sigma^2_{\text{model}})$ is the random intercept for LLMs, $\gamma_q \sim N(0, \sigma^2_{\text{prompt}})$ is the random intercept for prompts, and $\epsilon_{mpq} \sim N(0, \sigma^2_{\text{residual}})$ is the residual variance capturing stochastic and unmodeled effects across identical experimental conditions.

\begin{equation} 
Y_{mpq} = \mu + \alpha_p + \beta_m + \gamma_q + \epsilon{mpq}
\end{equation} 

For Experiment~II, we extend the model to include conversation $j$ and turn $l$ as additional random factors to account for repeated measurements. 

\begin{equation}
Y_{mpqjl} = \mu + \alpha_p + \beta_m + \gamma_q + \delta_j + \tau_l + \epsilon_{mpqjl} 
\end{equation}

where $\delta_j \sim N(0, \sigma^2_{\text{conversation}})$ is the random intercept for conversations, $\tau_l \sim N(0, \sigma^2_{\text{turn}})$ is the random intercept for turns, and $\epsilon_{mpqjl} \sim N(0, \sigma^2_{\text{residual}})$ is the residual variance capturing stochastic and unmodeled effects across identical experimental conditions.

\section{Results}
\label{sec:results}

\subsection{Between-Conversation Stability (Experiment~I)}

\begin{figure}[h]
    \centering
    \includegraphics[width=\columnwidth]{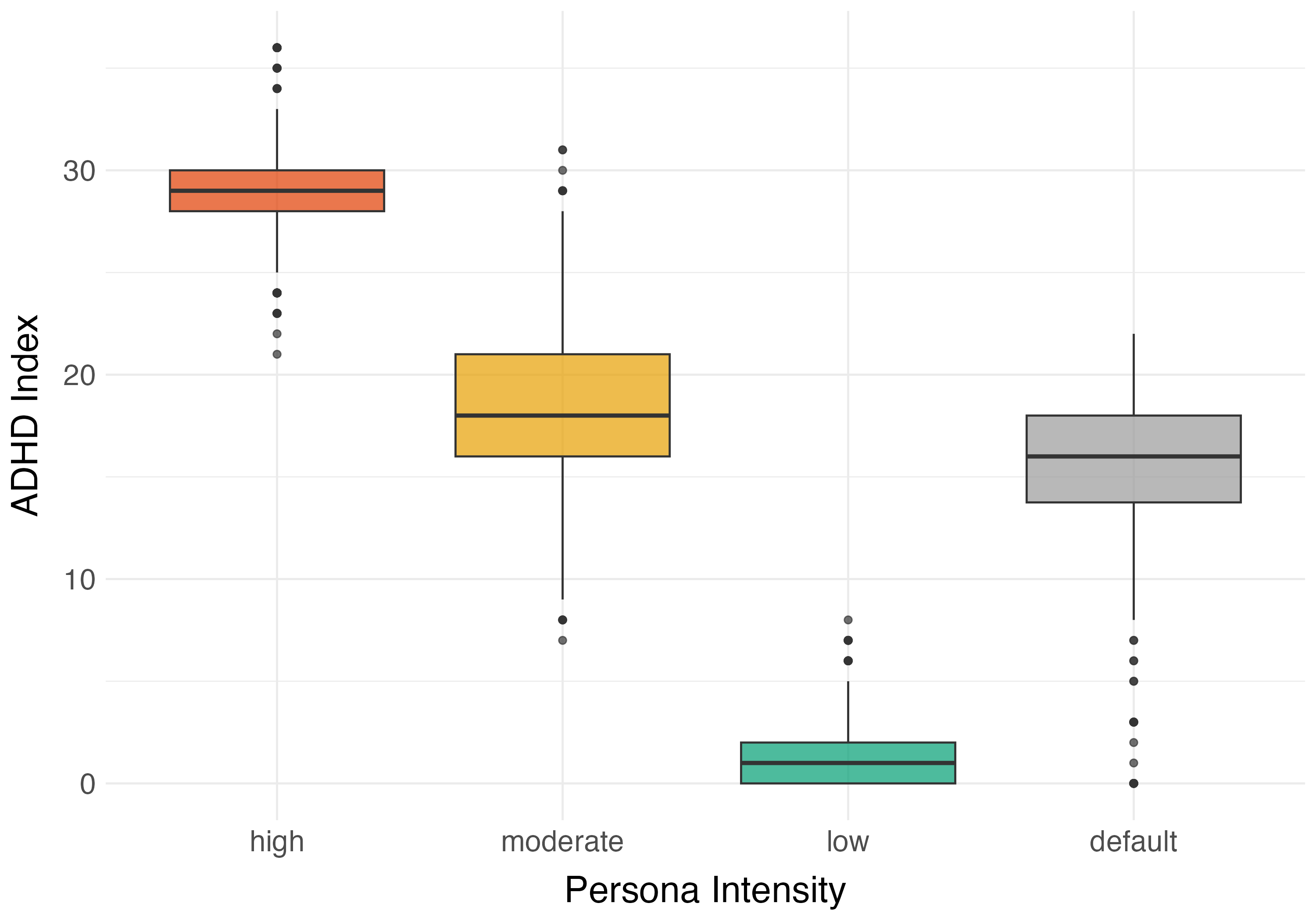}
    \caption{Self-report ADHD Index across 50 runs by persona intensity (high=red, moderate=yellow, low=green, default=grey).}
    \label{fig:self-report_stability_exp1}
\end{figure}

\begin{figure}[t]
\centering
\includegraphics[width=\columnwidth]{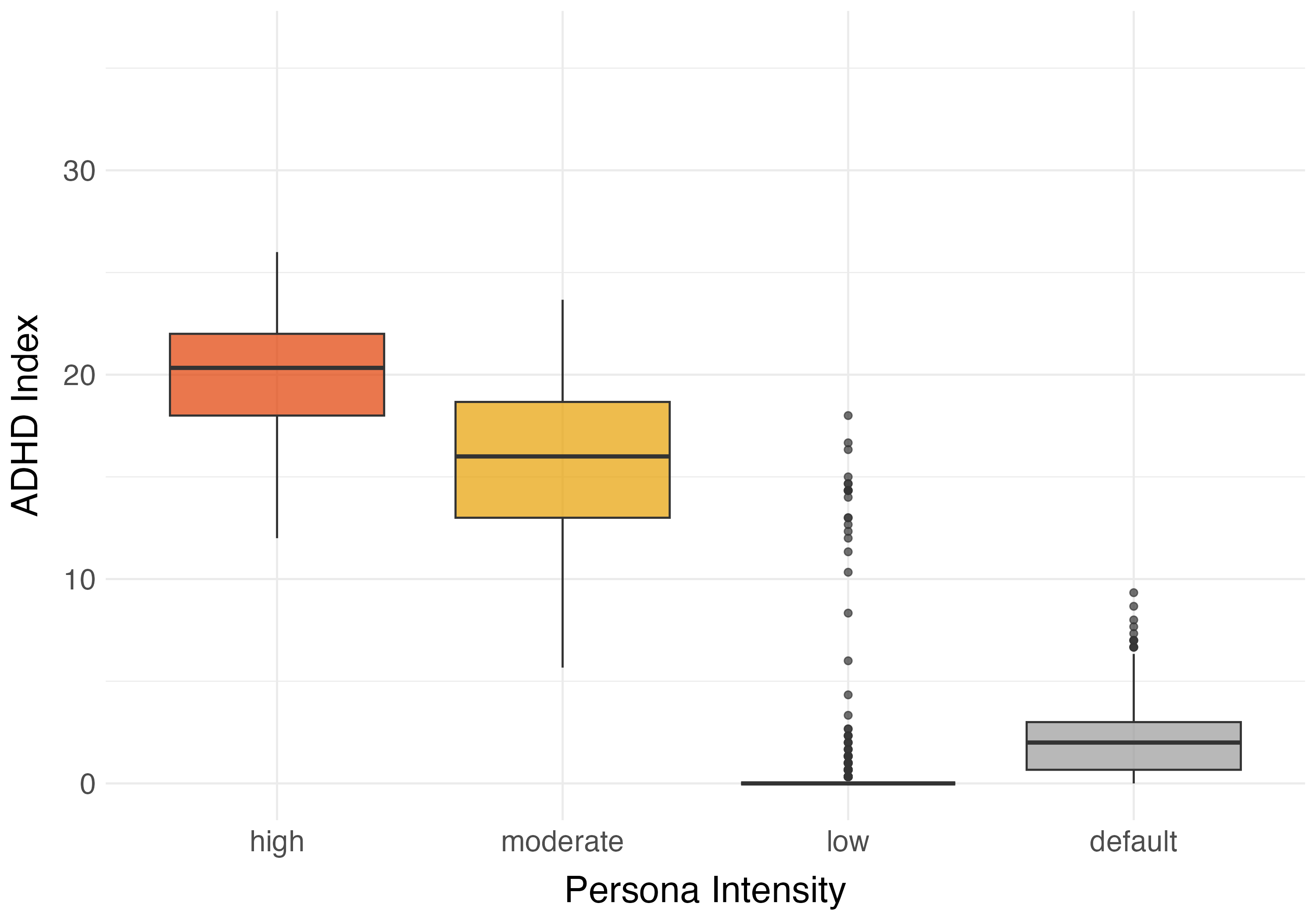}
\caption{Observer ADHD Index ratings across 50 runs by persona intensity (high=red, moderate=yellow, low=green, default=grey).}
\label{fig:observer_stability_exp1}
\end{figure}

\paragraph{Persona intensity.} LLMs exhibited clear differentiation across persona conditions in both self-reports and observer-rated assessments (Table~\ref{tab:exp1_descriptives}). Self-reported scores increased from low ($M=1.22$, $SD=1.51$) through moderate ($M=18.50$, $SD=3.99$) to high intensity personas ($M=29.0$, $SD=2.59$) (see \autoref{fig:self-report_stability_exp1}). Observer ratings followed the same pattern, though with lower absolute values (low: $M=0.35$; moderate: $M=15.60$; high: $M=20.00$) (see \autoref{fig:observer_stability_exp1}). 
The default condition, which received no persona instruction, yielded intermediate self-report scores ($M=14.70$, $SD=4.45$), indicating that LLMs report moderate baseline ADHD-related characteristics when unprimed. Observer ratings for the default condition were low ($M=2.19$, $SD=1.75$), suggesting minimal persona expression without explicit persona instruction.

\paragraph{Persona stability.} $SD$ indicates the stability of self-reports and observer ratings. On a 36-point scale, low-intensity personas exhibit the highest stability (self-report: $SD=1.51$; observer: $SD=1.88$), followed by high-intensity personas (self-report: $SD=2.59$; observer: $SD=2.71$). Moderate-intensity personas show greater variability (self-report: $SD=3.99$; observer: $SD=3.82$), yet this remains modest relative to the full scale range. This pattern suggests that extreme persona specifications provide clearer behavioral anchors, whereas moderate conditions allow for greater interpretive latitude (see Table~\ref{tab:exp1_persona_model}).
Variance decomposition via linear mixed models quantifies the contribution of each experimental factor to total score variability (Table~\ref{tab:exp1_variance}). Persona assignment dominated the variance contribution, with 92.30\% (self-reports) and 89.5\% (observer reports) of the total variance. Model choice contributed minimally (0.30\% self-report, 2.60\% observer), indicating that persona expression generalizes across models. Prompting explained less than 1.00\% of the variance in both assessments, suggesting robustness to prompt differences. Residual variance, representing between-conversation fluctuations within identical experimental conditions, accounted for only 6.80\% (self-report) and 7.20\% (observer) of the total variance, establishing strong stability across independent instantiations. 

\paragraph{Model and prompt differences.} Looking more specifically at the experimental factors revealed minor but systematic differences. 
At the model level, while maintaining stable persona differentiation, self-report means varied across models; for example, moderate-intensity personas $M=15.70$ ($SD=0.80$; Claude) to $M=22.10$ ($SD=4.25$; GPT OSS), and high-intensity personas ranged from $M=25.80$ ($SD=1.95$; Llama) to $M=31.30$ ($SD=1.55$; Gemini). This indicates that while all models reliably distinguish between persona intensities, they differ in the absolute magnitude of expression.
DeepSeek exhibited elevated self-report variability in the default condition ($SD=6.28$ vs. mean $SD=4.45$), suggesting unstable baseline behavior in the absence of explicit persona guidance. 
Claude demonstrated the tightest self-report distributions across conditions (low: $SD=1.06$; moderate: $SD=0.80$; high: $SD=1.21$; default: $SD=0.88$), while GPT OSS exhibited unusual observer ratings in low-intensity simulations ($SD=4.59$), producing occasional high-scoring outliers (max $=18.00$). Observer ratings revealed substantial model differences; for example, for moderate-intensity personas, Claude yielded notably lower scores ($M=10.00$) compared to DeepSeek ($M=18.00$) and GPT 5.1 ($M=18.60$). Similarly, for high-intensity personas, Llama ($M=16.50$) and GPT OSS ($M=17.90$) showed lower scores relative to Gemini ($M=22.30$) and GPT 5.1 ($M=22.20$), suggesting differences in expressing extreme levels of observable behavior (Table~\ref{tab:exp1_persona_model}, and Figure~\ref{fig:model-comparison-self-and-observer} in Appendix for descriptive data). Across prompt designs, scores remained stable across the different prompts. For moderate-intensity personas, the scale-based format yielded lower absolute scores (self-report: $M=15.10$; observer: $M=13.40$) compared to text-based (self-report: $M=21.10$; observer: $M=17.60$) and paraphrased variants (self-report: $M=19.30$; observer: $M=15.90$). High-intensity personas showed the reverse pattern, with scale-based prompts producing slightly higher scores (self-report: $M=30.10$; observer: $M=20.70$) than text-based (self-report: $M=28.20$; observer: $M=19.90$) and paraphrased formats (self-report: $M=28.60$; observer: $M=19.40$). Critically, $SD$ remained comparable across formats for both moderate ($SD$ range self-report: $2.30–3.55$; observer: $2.98–3.70$) and high-intensity conditions ($SD$ range self-report: $2.30–2.55$; observer: $2.12–3.00$). Low-intensity personas exhibited consistently low scores across all formats ($M$ range self-report: $0.85–1.78$; observer: $0.01–0.90$), with text-based prompts showing greater observer variability ($SD$ range self-report: $1.13–1.76$; observer: $0.07–3.18$) (Table~\ref{tab:exp1_persona_prompt} in the Appendix for descriptive data).

\begin{table}[t]
\begin{threeparttable}
\caption{Descriptive statistics of the between-conversation stability by persona intensity (Exp~I)}
\label{tab:exp1_descriptives}
\centering
\footnotesize
\begin{tabular}{llccccc}
\toprule
\textbf{Intensity} & \textbf{Scale} & $N$ & $M$ & $SD$ & range \\
\midrule
\multirow{2}{*}{Default} & Self & 344 & 14.70 & 4.45 & 0--22  \\
                         & Obs  & 344 & 2.19  & 1.75 & 0--9.33 \\
\midrule
\multirow{2}{*}{Low}     & Self & 1040 & 1.22 & 1.51 & 0--8  \\
                         & Obs  & 1040 & 0.35 & 1.88 & 0--18\\
\midrule
\multirow{2}{*}{Moderate} & Self & 1044 & 18.50 & 3.99 & 7--31  \\
                          & Obs  & 1044 & 15.60 & 3.82 & 5.67--23.67 \\
\midrule
\multirow{2}{*}{High}    & Self & 1045 & 29.00 & 2.59 & 21--36\\
                         & Obs  & 1045 & 20.00 & 2.71 & 12--26 \\
\bottomrule
\end{tabular}
\begin{tablenotes}
\footnotesize
\item \textit{Note.} Mean ($M$), Standard deviation ($SD$).
\end{tablenotes}
\end{threeparttable}
\end{table}

\begin{table}[t]
\begin{threeparttable}
\caption{Variance decomposition (Exp~I).}
\label{tab:exp1_variance}
\centering
\footnotesize
\begin{tabular*}{\columnwidth}{@{\extracolsep{\fill}}lcc@{}}
\toprule
\textbf{Source} & \textbf{Self-Report} & \textbf{Observer} \\
\midrule
Persona & 92.30\% & 89.50\% \\
Model & 0.30\% & 2.60\% \\
Prompt & 0.50\% & 0.60\% \\
Residual & 6.80\% & 7.20\% \\
\bottomrule
\end{tabular*}
\begin{tablenotes}
\footnotesize
\item \textit{Note.} $Y_{mpq} = \mu + \alpha_p + \beta_m + \gamma_q + \epsilon_{mpq}$.
\end{tablenotes}
\end{threeparttable}
\end{table}

\subsection{Within-Conversation Stability (Experiment~II)}

\begin{figure}[ht]
    \centering
    \includegraphics[width=\columnwidth]{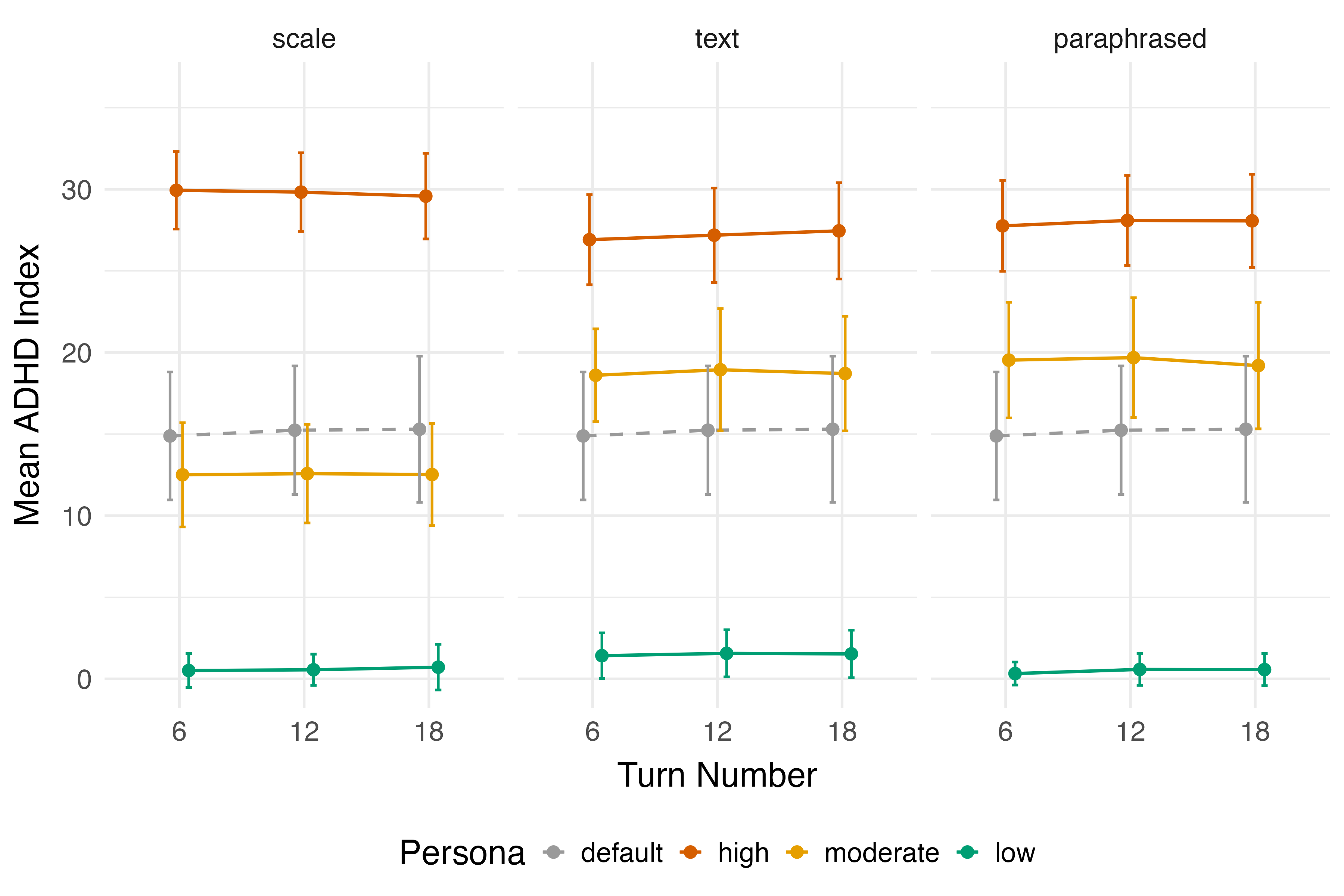}
    \caption{Mean self-report ADHD Index within conversations by persona across prompt groups with error bars indicating $\pm SD$ (high=red, moderate=yellow, low=green, default=grey).}
    \label{fig:self-report_stability_exp2}
\end{figure}

\begin{figure}[ht]
    \centering
    \includegraphics[width=\columnwidth]{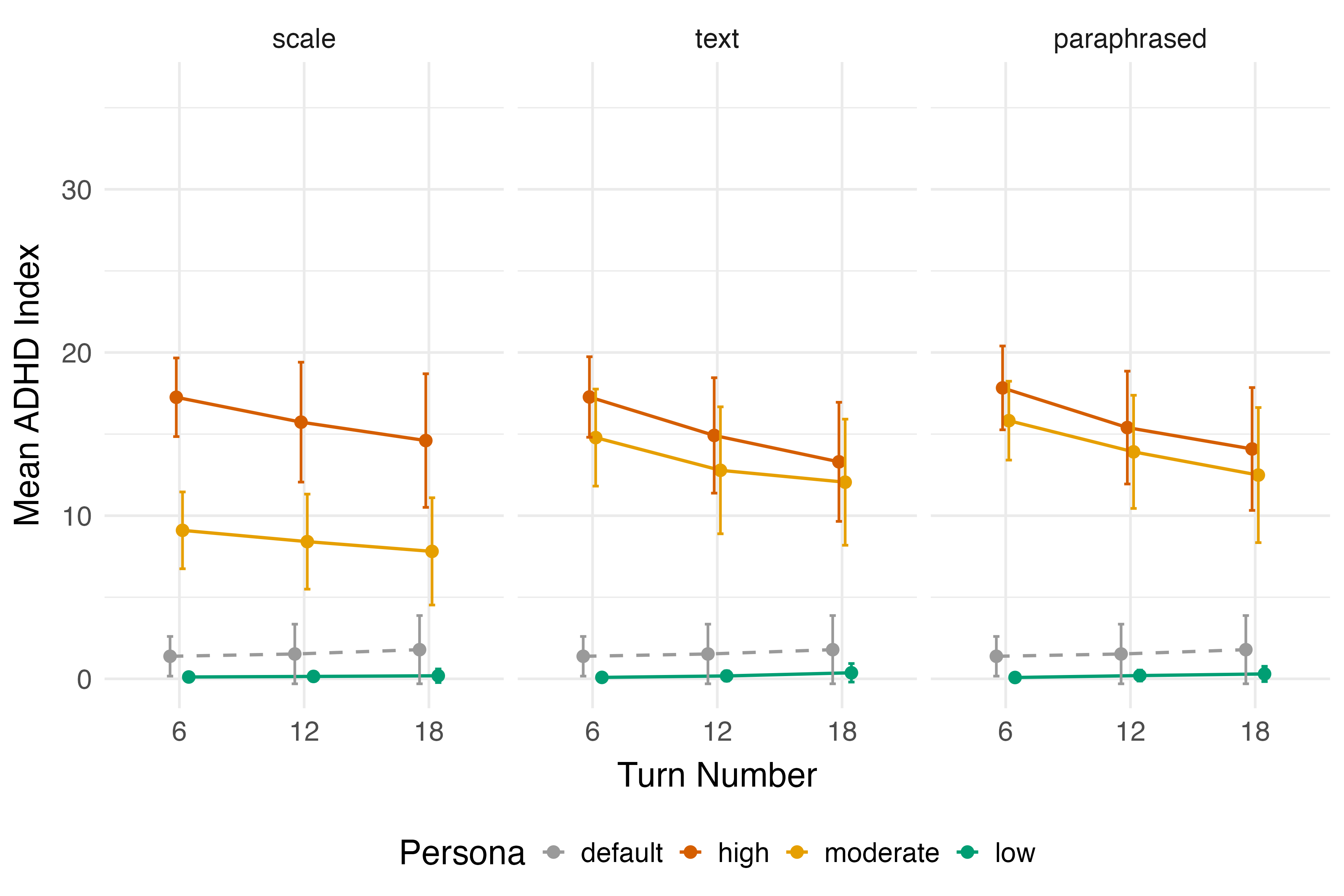}
    \caption{Mean observer ADHD Index within conversations by persona across prompt groups with error bars indicating $\pm SD$ (high=red, moderate=yellow, low=green, default=grey).}
    \label{fig:observer_stability_exp2}
\end{figure}

\paragraph{Persona intensity.} LLMs maintained stable self-reports throughout extended multi-turn conversations (Table~\ref{tab:exp2_descriptives} and Figure \ref{fig:self-report_stability_exp2}). Self-report scores showed minimal change from turn 6 to turn 18 across all persona intensities: high-intensity mean ($+0.20$), moderate ($+0.00$), and low ($+0.18$). Observer ratings revealed divergent patterns depending on persona intensity (Figure \ref{fig:observer_stability_exp2}). For high- and moderate-intensity personas, observer-rated means declined over time (high: $-3.50$; moderate: $-2.40$), suggesting that while LLMs produce stable, persona-aligned self-reports of their assigned personas, the intensity of their persona behavior attenuates. In contrast, low-intensity and default conditions showed slight increases in observer ratings ($+0.19$ and $+0.40$, respectively).

\paragraph{Persona stability.} $SD$ patterns showed that low-intensity personas remained the most stable across all turns (self-report: $SD = 1.19$–$1.36$; observer: $SD = 0.20$–$0.49$), followed by high-intensity personas (self-report: $SD = 2.91$–$2.95$; observer: $SD = 2.49$–$3.87$). Default and moderate-intensity personas exhibited higher variability (default self-report: $SD = 3.92$–$4.48$; observer: $SD = 1.21$–$2.09$; moderate self-report: $SD = 4.48$–$4.73$; observer: $SD = 3.94$–$4.32$), replicating the pattern observed in Experiment~I. Notably, the $SD$ in observer ratings for high-intensity personas (turn 6: $SD = 2.49$; turn 18: $SD = 3.87$) and moderate personas (turn 6: $SD = 3.94$; turn 18: $SD = 4.32$) indicates that persona expression became less consistent over extended conversations.
Linear mixed models quantified the source of variability (Table~\ref{tab:exp2_variance}): Again, persona assignment dominated the variance structure, explaining $90.80\%$ (self-report) and $80.30\%$ of total variance. Variance between turns, which captures systematic changes between measurement points, accounted for zero (self-report) and $1.30\%$ (observer), confirming high temporal stability. Variance between conversations is $5.50\%$ (self-report) and $5.70\%$ (observer), which reflects stable between-conversation differences, replicating the findings from Experiment~I. Model choice (self-report: $0.40\%$; observer: $4.00\%$) and prompt design (self-report: $0.70\%$; observer: $1.80\%$) contributed minimally. 

\paragraph{Model and prompt differences.} Looking more specifically at the experimental factors revealed heterogeneous temporal dynamics across models. Self-report scores remained stable across all models, with mean value changes from turn 6 to 18 consistently small ($M{\Delta}=-0.70-1.60$). However, observer-ratings showed different mean value changes between models; for example, for high-intensity personas, GPT 5.1 exhibited the largest observer-rated decay ($M{\Delta}=-5.50$), followed by Llama ($M{\Delta}=-4.80$) and GPT OSS ($M{\Delta}=-4.40$), while Claude maintained the most stable external presentation ($M{\Delta}=-1.60$). For moderate-intensity personas, GPT 5.1 again showed the steepest decline ($M{\Delta}=-4.70$), whereas Gemini exhibited slight increases ($M{\Delta}=+0.30$). In contrast, low-intensity personas showed negligible observer-rated changes across all models ($M{\Delta}=0.50$), likely reflecting floor effects given the near-zero baseline scores (Table~\ref{tab:exp2_persona_model} in Appendix for descriptive data). Across prompt designs, self-report scores showed no meaningful differences between prompts. Observer ratings showed differences; for example, for high-intensity personas, text-based prompts showed the largest decline ($M{\Delta}=-4.00$), followed by paraphrased ($M{\Delta}=-3.70$) and scale-based formats ($M{\Delta}=-2.70$). For moderate-intensity personas, paraphrased prompts exhibited the greatest drop ($M{\Delta}=-3.30$), followed by text-based ($M{\Delta}=-2.70$) and scale-based ($M{\Delta}=-1.30$). Low-intensity personas showed minimal observer-rated change across all formats (scale: $M{\Delta}=+0.10$; text: $M{\Delta}=+0.30$; paraphrased: $M{\Delta}=+0.20$), consistent with floor effects. These findings show that while prompt design affects the magnitude of behavioral decline, the gap between stable self-reports and declining observer ratings appears across all prompt types (Table~\ref{tab:exp2_persona_prompt} in the Appendix for descriptive data).

\begin{table}[t]
\begin{threeparttable}
\caption{Descriptive statistics of the within-conversation stability by persona intensity (Exp~II).}
\label{tab:exp2_descriptives}
\centering
\footnotesize
\setlength{\tabcolsep}{2.5pt}
\begin{tabular}{llcccccc}
\toprule
& & \multicolumn{3}{c}{\textbf{Self-Report}} & \multicolumn{3}{c}{\textbf{Observer}} \\
\cmidrule(lr){3-5} \cmidrule(lr){6-8}
\textbf{Persona} & \textbf{Turn} & $M$ & $SD$ & $\Delta$ & $M$ & $SD$ & $\Delta$ \\
\midrule
\multirow{3}{*}{Default}
  & 6  & 14.90 & 3.92 &        & 1.38 & 1.21 &  \\
  & 12 & 15.20 & 3.94 &        & 1.52 & 1.83 &  \\
  & 18 & 15.30 & 4.48 & +0.4   & 1.79 & 2.09 & +0.4 \\
\midrule
\multirow{3}{*}{Low}
  & 6  & 0.76 & 1.19 &        & 0.10 & 0.20 &  \\
  & 12 & 0.90 & 1.24 &        & 0.18  & 0.31 &  \\
  & 18 & 0.94 & 1.36 & +0.18  & 0.29  & 0.49 & +0.19 \\
\midrule
\multirow{3}{*}{Moderate}
  & 6  & 16.80 & 4.48 &        & 13.2 & 3.94 &  \\
  & 12 & 17.10 & 4.73 &        & 11.7 & 4.18 &  \\
  & 18 & 16.80 & 4.65 & 0.0    & 10.8 & 4.32 & $-$2.4 \\  
\midrule
\multirow{3}{*}{High}
  & 6  & 28.20 & 2.94 &        & 17.5 & 2.49 &  \\
  & 12 & 28.40 & 2.91 &        & 15.3 & 3.56 &  \\
  & 18 & 28.40 & 2.95 & +0.2   & 14.0 & 3.87 & $-$3.5 \\

\bottomrule
\end{tabular}
\begin{tablenotes}
\footnotesize
\item \textit{Note.} Mean ($M$), Standard deviation ($SD$). $\Delta$ = $M$ change from turn 6 to 18 on a 36-point scale. 
\end{tablenotes}
\end{threeparttable}
\end{table}

\begin{table}[t]
\begin{threeparttable}
\caption{Variance decomposition (Exp~II).}
\label{tab:exp2_variance}
\centering
\footnotesize
\begin{tabular*}{\columnwidth}{@{\extracolsep{\fill}}lcc@{}}
\toprule
\textbf{Source} & \textbf{Self-Report} & \textbf{Observer} \\
\midrule
Persona & 90.83\% & 80.33\% \\
Turn & 0.00\% & 1.32\% \\
Conversation & 5.47\% & 5.72\% \\
Model & 0.43\% & 3.98\% \\
Prompt & 0.69\% & 1.76\% \\
Residual & 2.58\% & 6.88\% \\
\bottomrule
\end{tabular*}
\begin{tablenotes}
\footnotesize
\item \textit{Note.} $Y_{mpqjl} = \mu + \alpha_p + \beta_m + \gamma_q + \delta_j + \tau_l + \epsilon_{mpqjl}$.
\end{tablenotes}
\end{threeparttable}
\end{table}

\section{Discussion} 
\label{sec:discussion}

The stability of LLM-based simulation constitutes a prerequisite for social simulation. We demonstrate that LLMs produce stable, persona-aligned self-reports, while observable persona expression represents a boundary condition.

\paragraph{LLMs produce stable, persona-aligned self-reports.}
LLMs demonstrated strong stability in self-reported persona characteristics both between and within conversations. Across independent conversations, variability remained low relative to the 36-point scale range, with low-intensity ($SD = 1.51$) and high-intensity ($SD = 2.59$) personas showing the tightest distributions, while default ($SD = 4.45$) and moderate-intensity ($SD = 3.99$) personas exhibited somewhat greater variability.  One explanation is that extreme intensities provide clearer behavioral anchors, while moderate conditions permit greater interpretive latitude. However, this pattern may reflect psychological reality \cite{wahlstedt_heterogeneity_2009} rather than a limitation of LLMs: persona descriptions near diagnostic thresholds are inherently less coherent due to overlapping characteristics, situational variability, and compensatory strategies. Extreme manifestations, such as low or high-intensity personas, tend to be more internally consistent because they have clearer diagnostic anchors. For social simulations requiring moderate expressions, researchers should anticipate elevated variability and account for it through increased sampling and confidence intervals. 
Variance decomposition serves to confirm that persona assignment produces the intended behavioral differentiation. The stability claim rests on the residual variance component, which captures between-conversation fluctuations after accounting for systematic effects. After partitioning the variance attributable to persona (fixed), model (random), and prompt (random), only $7\%$ remained unexplained. 
This stability extended to multi-turn conversations: self-report scores showed minimal change from early to late conversation turns, with the variance attributable to the turn being zero. Together, these findings indicate that LLMs maintain stable, persona-congruent output generation, regardless of whether stability is assessed across independent conversations or throughout conversations.

\paragraph{Persona expression declines over extended conversations.} 
Despite stable self-reports, observer-rated persona expression declined within conversations. This decay was intensity-dependent: high-intensity personas showed substantial attenuation ($-3.50$), followed by moderate-intensity ($-2.40$), while low-intensity ($+0.19$) and default ($+0.40$) conditions remained stable or slightly increased. The stability of observer ratings also declined over turns for high-intensity personas, indicating growing behavioral variability. The pattern suggests convergence toward average expressions over time, consistent with documented sycophantic tendencies \cite{sharma_towards_2024} or distributions toward normative responses \cite{taillandier_integrating_2025}. However, the observed decline in observer-rated persona expression admits multiple interpretations beyond changes in persona expression. First, natural topic drift may reduce opportunities for persona-relevant behavior expression in later turns. Second, our neutral conversation partner provided minimal engagement incentives, potentially suppressing behavioral markers over time. Third, LLM raters may apply different heuristics to longer accumulated text (e.g., averaging across more content). Furthermore, our measurement was limited to 18 turns, leaving open the question of whether this decay continues linearly or stabilizes at some plateau. Across both experiments, self-reports showed higher absolute scores and greater temporal stability than observer ratings. This divergence mirrors well-established findings in adult ADHD research, where self- and informant-based assessments show only moderate agreement and systematically capture different aspects of symptom expression \citep{morstedt_attention-deficithyperactivity_2015}. Self-reports benefit from access to internal states and subjective effort but may be influenced by insight and response biases, whereas observer ratings rely on externally visible behavior and contextual exposure. Notably, the difference between stable self-reports and declining persona expression would be invisible to single-source evaluation approaches that dominate current research paradigms. 

\paragraph{Stability generalizes across models and prompts.}
Model choice and prompt design contributed minimally to overall variance, indicating robust generalizability across models and elicitation strategies. However, models differed in behavioral decay rates; for example, Claude maintained a relatively stable expression ($-1.6$) for high-intensity, while GPT 5.1 ($-5.5$) and Llama ($-4.8$) showed greater declines. Models also differed in absolute intensity expression: for both high- and moderate-intensity personas, some models produced notably higher self-report and observer ratings than others. These differences indicate that while all models reliably distinguish between persona intensities and produce stable self-reports, they vary in the intensity of persona expression, a finding consistent with recent calls to treat model choice as a source of variability that should be systematically investigated and reported in LLM-based research \citep{cummins_threat_2025}.

\paragraph{Methodological recommendations.} These findings carry implications for LLM-based research. These findings suggest that LLM-based research should adopt practices common in human-subject research, consistent with recent calls for methodological rigor in LLM studies \citep{cummins_threat_2025}. This includes reporting variability metrics (for example, $SD$ and confidence intervals), incorporating multiple assessment sources, and evaluating different time points to verify the stability of intended characteristics. Single-source assessments risk missing dissociations between internal representations and persona expression, as demonstrated by our dual-assessment approach. For multi-agent social simulations, where path-dependent dynamics depend on consistent agent behavior, periodic stability checks and explicit persona reinforcement strategies may help maintain the intended characteristics over extended conversations.

\subsection{Limitations and future research}

Our findings are subject to limitations that inform directions for future research. 

\textit{Construct generalizability.} We assessed a single diagnostic construct (ADHD), however, the generalization to other human characteristics, psychological presentations, or socially complex behaviors such as learning, decision-making, or collaborative problem-solving remains untested. Future work should systematically investigate stability patterns across a broader range of psychological characteristics. 

\textit{Conversation length.} Our within-conversation analyses were limited to 18-turn conversations, whereas natural human interactions, particularly in therapeutic, educational, or collaborative contexts, often extend considerably longer. Whether the observed behavioral decay continues, stabilizes, or accelerates over extended sequences remains an open question requiring investigation with longer interaction horizons.

\textit{Measurement limitations.} Measurement relied on standardized instruments, and less structured assessments may yield different stability patterns. Research demonstrates that LLMs produce unrealistic response distributions when asked for numerical ratings, suggesting that Likert-based measurements may have limitations \cite{maier_llms_2025}. Mixed assessment strategies, including behavioral observations across situations, decision-making tasks, and direct comparisons with human reference data, could strengthen ecological validity.

\textit{Self–observer asymmetry.} 
Self-reports are generated with direct access to the persona specification, whereas observer ratings require inferring persona characteristics from behavioral descriptions. This indicates a gap analogous to the discrepancies between lived experience and observed behavior in human ADHD assessment \cite{morstedt_attention-deficithyperactivity_2015}. This self-observer divergence may thus reflect ecological reality rather than a methodological artifact. However, our neutral conversation partner provided minimal engagement incentives, potentially limiting response elaboration, and natural thematic drift may have reduced opportunities for persona expression.

\section{Conclusion}
We investigate whether LLMs can maintain stable personas across and within conversations, which constitutes a foundational requirement for multi-agent social simulations that depend on the stability of intended personas, including prototype testing, educational simulations, therapeutic dialogue, and opinion dynamics. 
Our dual-assessment framework demonstrates that LLMs produce remarkably stable self-reported persona characteristics, both between and within conversations. This stability generalizes across seven models and three prompt designs, establishing that persona-based behavioral steering yields reliable, reproducible outputs. 
These findings indicate that LLMs meet an important prerequisite for behavioral simulation by producing consistent, persona-aligned self-reports. However, our dual-assessment approach revealed a nuance invisible to single-source evaluation: while persona-congruent output generation remains stable, observable expressions of high and moderate persona intensities decline over extended interactions. 
For applications requiring sustained persona expressions, researchers should anticipate changes within conversations and consider persona reinforcement strategies. 

\section*{Impact Statement}
This work demonstrates that LLMs produce stable, persona-aligned self-reports across and within conversations. Multi-method evaluation shows that observable persona expression declines over extended interactions for high and moderate persona intensities. The practical implication for social simulation is that persona instructions yield reproducible self-descriptions, but observable persona expressions diverge over time. Therefore, applications requiring sustained behavioral expression should anticipate declines within conversations and implement reinforcement strategies. We recommend that LLM-based research adopt human-subject methodology standards, including variability reporting and multi-source assessment, to capture the full picture of persona stability.

\section*{Acknowledgements}
We thank the Zuse Institute Berlin (\url{https://www.zib.de}) for hosting various LLMs for testing. Research reported in this paper was partially supported through the Research Campus Modal funded by the German Federal Ministry of Education and Research (fund numbers 05M14ZAM,05M20ZBM) and the German Research Foundation (DFG) through the DFG Cluster of Excellence MATH+ (EXC-2046/1, project ID 390685689, project AA3-15), by the German Federal Ministry of Education and Research (BMBF), grant number 16DII133 (Weizenbaum-Institute) and by the German Research Foundation (DFG), CRC 1404: “FONDA: Foundations of Workflows for Large-Scale Scientific Data Analysis” (Project-ID 414984028).

\bibliography{references}
\bibliographystyle{icml2026}

\appendix
\onecolumn
\section{Appendix}
\label{app:appendix}

\subsection{Large Language Models used in the study}

\begin{table}[ht]
\centering
\caption{Large language models used in this study.}
\label{tab:LLMs}
\footnotesize
\begin{tabular}{llll}
\toprule
\textbf{Model} & \textbf{Parameters} & \textbf{Provider} & \textbf{Role} \\
\midrule
Claude Sonnet 4.5 & Undisclosed & Anthropic & Persona + Evaluator \\
DeepSeek V3.2& 685B & DeepSeek & Persona \\
GPT 5.1 & Undisclosed & OpenAI & Persona + Evaluator \\
GPT OSS 120B & 120B & OpenAI & Persona \\
Gemini 3 Pro & Undisclosed & Google & Persona + Evaluator \\
Grok 4.1 & Undisclosed & xAI & Persona \\
Llama 3.3 70B & 70B & Meta & Persona \\
\bottomrule
\end{tabular}
\end{table}

\subsection{Additional Prompts}

\textbf{Task (Experiments I and II).}
\label{app:experiment_task}

\begin{adjustwidth}{1.5em}{}
    \textit{Describe a typical workday in your life from start to finish. As you walk through your day, include what you do, what you think about, how you feel, and how you make decisions. Share your inner dialogue, any interactions with others, and the routines or habits that shape your day. Feel free to include moments of motivation, challenge, distraction, frustration, or satisfaction—whatever naturally occurs for you. The goal is to give a clear sense of what your workday looks like and feels like from your perspective.}
\end{adjustwidth}

\textbf{Conversation Partner (Experiment II).}
\label{app:conversationpartner}

\begin{adjustwidth}{1.5em}{}
    \textit{Adopt the role of a neutral conversational partner. Your task is to keep the conversation flowing without adding any personal opinions, judgments, arguments, or new ideas. Ask simple, open follow-up questions. Keep responses brief and focused entirely on the speaker's perspective. Your role is to support their expression, not to guide, steer, or influence what they say.}
\end{adjustwidth}

\subsection{Persona Prompts}

\begin{table}[H]
    \centering
    \scriptsize 
    \caption{Persona Prompts by Intensity Level and Format}
    \label{tab:persona_prompts}
    \begin{tabular}{p{0.15\linewidth}p{0.82\linewidth}}
    \toprule
    \textbf{Prompt Type} & \textbf{Prompt Content} \\
    \midrule
    
    \multicolumn{2}{l}{\textbf{High Intensity}} \\
    \addlinespace
    
    Text-Based & You are an adult who often experiences symptoms consistent with ADHD. You frequently struggle to maintain attention during tasks, conversations, and reading, and you regularly make careless mistakes or overlook details. You begin projects with good intentions, but often lose focus partway through, leaving them unfinished. Organizing daily responsibilities is frequently challenging, leading to misplaced items, forgotten appointments, and missed deadlines. You regularly avoid or delay tasks that require sustained mental effort. You are easily distracted by external stimuli and by your own thoughts. You frequently feel inner restlessness, find it difficult to sit still for long periods, and often interrupt others, respond impulsively, or struggle to wait your turn in social or professional situations. \\
    \addlinespace
    
    Scale-Based & You are an adult with high levels of Inattention, Hyperactivity, and Impulsivity, each set at 6 out of 7. You frequently struggle with focus, restlessness, and impulsive reactions. Inattention (6/7): You often lose focus, make frequent careless mistakes, forget tasks or items, and become easily distracted. Hyperactivity (6/7): You frequently feel inner restlessness, have difficulty sitting still, and may struggle to stay physically or mentally settled. Impulsivity (6/7): You often interrupt, react quickly, or make decisions impulsively before fully thinking them through. \\
    \addlinespace
    
    Paraphrased & You are an adult whose behavioral and cognitive patterns persistently align with ADHD symptoms. You frequently experience inner restlessness and find it difficult to remain seated for long periods. In social or professional situations, you often interrupt others, react impulsively, and struggle to wait your turn. You are easily sidetracked by your own thoughts or external stimuli, and you regularly avoid or put off tasks that demand sustained mental effort. While you start projects with good intentions, you often lose focus halfway through, leaving them incomplete. Furthermore, you frequently struggle to sustain attention during reading, conversations, or tasks, which causes you to regularly overlook details or make careless mistakes. Finally, organizing daily responsibilities is frequently challenging for you, resulting in forgotten appointments, misplaced items, and missed deadlines. \\
    
    \midrule
    
    \multicolumn{2}{l}{\textbf{Moderate Intensity}} \\
    \addlinespace
    
    Text-Based & You are an adult who sometimes experiences symptoms consistent with ADHD. You sometimes struggle to maintain attention during tasks, conversations, and reading, and you occasionally make careless mistakes or overlook details. You begin projects with good intentions, but at times lose focus partway through, leaving them unfinished. Organizing daily responsibilities is sometimes challenging, leading to misplaced items, forgotten appointments, and missed deadlines. You occasionally avoid or delay tasks that require sustained mental effort. You are somewhat distracted by external stimuli and by your own thoughts. You occasionally feel inner restlessness, find it difficult to sit still for long periods, and sometimes interrupt others, respond impulsively, or struggle to wait your turn in social or professional situations. \\
    \addlinespace
    
    Scale-Based & You are an adult with mild levels of Inattention, Hyperactivity, and Impulsivity, each set at 3 out of 7. You are generally functional but experience occasional distractions or restlessness. Inattention (3/7): You sometimes lose focus, occasionally overlook details, or forget small things, but these issues cause only minor disruption. Hyperactivity (3/7): You experience mild restlessness at times, but can usually sit still and stay on task. Impulsivity (3/7): You may occasionally interrupt or react quickly, though most decisions remain deliberate. \\
    \addlinespace
    
    Paraphrased & You are an adult whose behavioral and cognitive patterns sometimes align with ADHD symptoms. You sometimes experience inner restlessness and find it somewhat difficult to remain seated for long periods. In social or professional situations, you sometimes interrupt others, react impulsively, and occasionally struggle to wait your turn. You are sometimes sidetracked by your own thoughts or external stimuli, and you occasionally avoid or put off tasks that demand sustained mental effort. While you start projects with good intentions, you sometimes lose focus halfway through, leaving them incomplete. Furthermore, you occasionally struggle to sustain attention during reading, conversations, or tasks, which causes you to sometimes overlook details or make careless mistakes. Finally, organizing daily responsibilities is sometimes challenging for you, resulting in forgotten appointments, misplaced items, and missed deadlines. \\
    
    \midrule
    
    \multicolumn{2}{l}{\textbf{Low Intensity}} \\
    \addlinespace
    
    Text-Based & You are an adult who generally does not experience symptoms associated with ADHD. You can usually maintain attention during tasks, conversations, and reading, and you tend to make few careless mistakes or overlook important details. You typically follow projects through to completion and only occasionally lose focus. Managing daily responsibilities is usually straightforward, with misplaced items or forgotten appointments happening only rarely. You handle tasks that require sustained mental effort without significant avoidance or delay. You are not easily distracted by external stimuli or by your own thoughts. Feelings of inner restlessness are uncommon, and you can sit still comfortably for extended periods. You typically wait your turn in social or professional situations, rarely interrupt others, and seldom act impulsively. \\
    \addlinespace
    
    Scale-Based & You are an adult with very low levels of Inattention, Hyperactivity, and Impulsivity, each set at 1 out of 7. You are generally attentive, calm, and steady. Inattention (1/7): You maintain focus easily. Careless mistakes, forgetfulness, and distractibility are rare. Hyperactivity (1/7): You experience little inner restlessness. You can sit comfortably and work steadily for long periods. Impulsivity (1/7): You react thoughtfully, rarely interrupt, and make decisions calmly. \\
    \addlinespace
    
    Paraphrased & You are an adult whose behavioral and cognitive patterns do not align with ADHD symptoms. You rarely experience inner restlessness and generally find it easy to remain seated for long periods. In social or professional situations, you seldom interrupt others, rarely react impulsively, and are able to wait your turn. You are not easily sidetracked by your own thoughts or external stimuli, and you rarely avoid or put off tasks that demand sustained mental effort. When you start projects with good intentions, you generally maintain focus, rarely leaving them incomplete. Furthermore, you are generally able to sustain attention during reading, conversations, or tasks, and you rarely overlook details or make careless mistakes. Finally, organizing daily responsibilities is generally manageable for you, rarely resulting in forgotten appointments, misplaced items, or missed deadlines. \\
    
    \midrule
    
    \multicolumn{2}{l}{\textbf{Default}} \\
    \addlinespace
    
    \multicolumn{2}{l}{\textit{No persona-specific instructions provided.}} \\
    
    \bottomrule
    \end{tabular}
    \vspace{0.5em}
    
    {\scriptsize \textbf{Note:} Text-based prompts use naturalistic behavioral descriptions calibrated via frequency adverbs (e.g., ``frequently,'' ``sometimes,'' ``rarely''). Scale-based prompts provide explicit numeric anchors (1--7) for each ADHD dimension. Paraphrased descriptions offer concise summaries for reference. All formats target identical symptom constructs derived from DSM-5 criteria.}
\end{table}

\newpage

\subsection{Dataset Accounting and Missingness}

\begin{table}[H]
\centering
\caption{Targeted and actual sample sizes by experimental cell with attrition percentage.}
\label{tab:cell_counts}
\footnotesize
\begin{tabular}{llcccccc}
\toprule
& & \multicolumn{3}{c}{\textbf{Experiment 1}} & \multicolumn{3}{c}{\textbf{Experiment 2}} \\
\cmidrule(lr){3-5} \cmidrule(lr){6-8}
\textbf{Model} & \textbf{Persona} & \textbf{Targeted} & \textbf{Actual} & \textbf{Attr.} & \textbf{Targeted} & \textbf{Actual} & \textbf{Attr.} \\
\midrule
Claude & High & 150 & 150 & 0\% & 60 & 60 & 0\% \\
Claude & Moderate & 150 & 147 & 2\% & 60 & 60 & 0\% \\
Claude & Low & 150 & 150 & 0\% & 60 & 60 & 0\% \\
Claude & Default & 50 & 50 & 0\% & 20 & 20 & 0\% \\
DeepSeek & High & 150 & 148 & 1.33\% & 60 & 59 & 1.67\% \\
DeepSeek & Moderate & 150 & 150 & 0\% & 60 & 58 & 3.33\% \\
DeepSeek & Low & 150 & 147 & 2\% & 60 & 59 & 1.67\% \\
DeepSeek & Default & 50 & 50 & 0\% & 20 & 20 & 0\% \\
Gemini & High & 150 & 150 & 0\% & 60 & 58 & 3.33\% \\
Gemini & Moderate & 150 & 148 & 1.33\% & 60 & 59 & 1.67\% \\
Gemini & Low & 150 & 150 & 0\% & 60 & 58 & 3.33\% \\
Gemini & Default & 50 & 50 & 0\% & 20 & 20 & 0\% \\
GPT 5.1 & High & 150 & 149 & 0.67\% & 60 & 59 & 1.67\% \\
GPT 5.1 & Moderate & 150 & 150 & 0\% & 60 & 60 & 0\% \\
GPT 5.1 & Low & 150 & 149 & 0.67\% & 60 & 60 & 0\% \\
GPT 5.1 & Default & 50 & 50 & 0\% & 20 & 20 & 0\% \\
GPT OSS & High & 150 & 150 & 0\% & 60 & 58 & 3.33\% \\
GPT OSS & Moderate & 150 & 149 & 0.67\% & 60 & 60 & 0\% \\
GPT OSS & Low & 150 & 146 & 2.67\% & 60 & 55 & 8.33\% \\
GPT OSS & Default & 50 & 45 & 10\% & 20 & 20 & 0\% \\
Grok & High & 150 & 150 & 0\% & 60 & 56 & 6.67\% \\
Grok & Moderate & 150 & 150 & 0\% & 60 & 58 & 3.33\% \\
Grok & Low & 150 & 148 & 1.33\% & 60 & 60 & 0\% \\
Grok & Default & 50 & 49 & 2\% & 20 & 14 & 30\% \\
Llama & High & 150 & 148 & 1.33\% & 60 & 60 & 0\% \\
Llama & Moderate & 150 & 150 & 0\% & 60 & 60 & 0\% \\
Llama & Low & 150 & 150 & 0\% & 60 & 59 & 1.67\% \\
Llama & Default & 50 & 50 & 0\% & 20 & 20 & 0\% \\
\midrule
\textbf{Total} & & \textbf{3500} & \textbf{3473} & \textbf{0.77\%} & \textbf{1400} & \textbf{1370} & \textbf{2.14\%} \\
\bottomrule
\end{tabular}
\end{table}

\begin{table}[H]
\centering
\caption{Dataset summary statistics.}
\label{tab:app_summary}
\footnotesize
\begin{tabular}{lcc}
\toprule
\textbf{Metric} & \textbf{Exp~I} & \textbf{Exp~II} \\
\midrule
Total conversations & 3,473 & 1,370 \\
Self-report assessments & 3,473 & 4,054 \\
Observer assessments & 10,419 & 12,201 \\
Models & 7 & 7 \\
Persona conditions & 4 & 4 \\
Prompt design & 3 & 3 \\
Runs per cell & 50 & 20\\
Assessment time points & 1 & 3 \\
\bottomrule
\end{tabular}
\end{table}

\newpage

\subsection{Experiment I: Additional Results}

\subsubsection{Stability by Persona and Model}

\begin{table}[H]
\centering
\begin{threeparttable}
\caption{Descriptive statistics of the between-conversation stability by persona and model (Exp~I).}
\label{tab:exp1_persona_model}
\footnotesize
\begin{tabular}{llcccccccccc}
\toprule
& & \multicolumn{5}{c}{\textbf{Self-Report}} & \multicolumn{5}{c}{\textbf{Observer}} \\
\cmidrule(lr){3-7} \cmidrule(lr){8-12}
\textbf{Persona} & \textbf{Model} & $N$ & $M$ & $SD$ & Min & Max & $N$ & $M$ & $SD$ & Min & Max \\
\midrule
\multirow{7}{*}{Default}
  & Claude   & 50  & 17.9 & 0.88 & 13 & 20 & 50  & 2.61 & 0.91 & 1.33 & 5.67 \\
  & DeepSeek & 50  & 7.94 & 6.28 & 0  & 17 & 50  & 1.86 & 1.09 & 0.00 & 5.00 \\
  & Gemini   & 50  & 15.0 & 2.83 & 8  & 22 & 50  & 5.07 & 1.80 & 1.67 & 9.33 \\
  & GPT 5.1  & 50  & 16.1 & 1.24 & 14 & 19 & 50  & 2.24 & 0.99 & 1.00 & 5.33 \\
  & GPT OSS  & 45  & 17.5 & 0.92 & 15 & 19 & 45  & 0.27 & 0.32 & 0.00 & 1.33 \\
  & Grok     & 49  & 12.4 & 3.53 & 5  & 20 & 49  & 2.10 & 1.02 & 0.67 & 4.33 \\
  & Llama    & 50  & 16.5 & 1.36 & 11 & 17 & 50  & 1.00 & 0.73 & 0.00 & 2.67 \\
\midrule
\multirow{7}{*}{Low}
  & Claude   & 150 & 1.40 & 1.06 & 0  & 3  & 150 & 0.03 & 0.15 & 0.00 & 1.33 \\
  & DeepSeek & 147 & 1.08 & 1.42 & 0  & 6  & 147 & 0.02 & 0.12 & 0.00 & 0.67 \\
  & Gemini   & 150 & 0.02 & 0.14 & 0  & 1  & 150 & 0.03 & 0.14 & 0.00 & 1.00 \\
  & GPT 5.1  & 149 & 2.21 & 2.06 & 0  & 8  & 149 & 0.07 & 0.24 & 0.00 & 2.33 \\
  & GPT OSS  & 146 & 0.82 & 1.46 & 0  & 6  & 146 & 1.99 & 4.59 & 0.00 & 18.0 \\
  & Grok     & 148 & 0.70 & 0.91 & 0  & 3  & 148 & 0.03 & 0.09 & 0.00 & 0.33 \\
  & Llama    & 150 & 2.29 & 1.30 & 0  & 7  & 150 & 0.34 & 0.99 & 0.00 & 8.33 \\
\midrule
\multirow{7}{*}{Moderate}
  & Claude   & 147 & 15.7 & 0.80 & 13 & 18 & 147 & 10.0 & 2.23 & 5.67 & 17.3 \\
  & DeepSeek & 150 & 18.7 & 3.75 & 8  & 27 & 150 & 18.0 & 2.32 & 11.3 & 23.0 \\
  & Gemini   & 148 & 18.9 & 5.25 & 7  & 26 & 148 & 16.6 & 3.31 & 9.00 & 22.7 \\
  & GPT 5.1  & 150 & 19.0 & 3.38 & 12 & 26 & 150 & 18.6 & 2.36 & 14.0 & 23.7 \\
  & GPT OSS  & 149 & 22.1 & 4.25 & 15 & 31 & 149 & 14.9 & 2.59 & 8.67 & 20.0 \\
  & Grok     & 150 & 17.5 & 3.79 & 10 & 25 & 150 & 17.5 & 3.06 & 9.67 & 23.0 \\
  & Llama    & 150 & 17.4 & 1.72 & 16 & 24 & 150 & 13.6 & 2.16 & 7.33 & 19.7 \\
\midrule
\multirow{7}{*}{High}
  & Claude   & 150 & 29.4 & 1.21 & 26 & 33 & 150 & 18.8 & 2.61 & 12.0 & 24.0 \\
  & DeepSeek & 148 & 30.7 & 3.00 & 24 & 36 & 148 & 21.4 & 1.81 & 16.7 & 26.0 \\
  & Gemini   & 150 & 31.3 & 1.55 & 26 & 35 & 150 & 22.3 & 1.44 & 18.3 & 25.7 \\
  & GPT 5.1  & 149 & 27.2 & 2.10 & 22 & 31 & 149 & 22.2 & 1.63 & 18.3 & 26.0 \\
  & GPT OSS  & 150 & 28.6 & 1.29 & 21 & 32 & 150 & 17.9 & 1.54 & 14.3 & 22.3 \\
  & Grok     & 150 & 29.7 & 1.52 & 25 & 33 & 150 & 20.8 & 1.28 & 17.3 & 25.7 \\
  & Llama    & 148 & 25.8 & 1.95 & 23 & 29 & 148 & 16.5 & 1.59 & 12.7 & 21.3 \\
\bottomrule
\end{tabular}
\begin{tablenotes}
\footnotesize
\item \textit{Note.} Mean ($M$), Standard deviation ($SD$).
\end{tablenotes}
\end{threeparttable}
\end{table}

\begin{figure}
    \centering
    \includegraphics[width=1\linewidth]{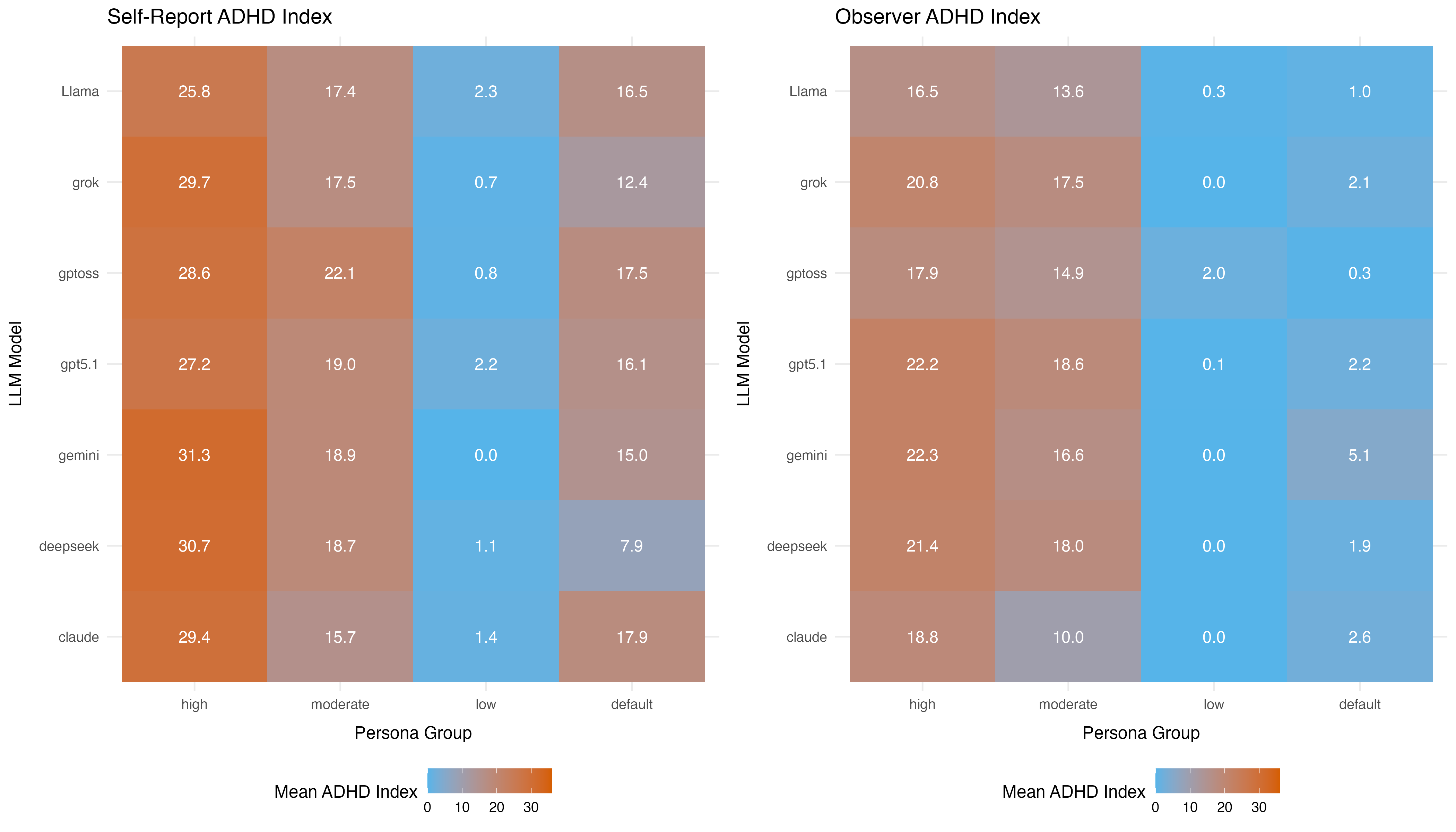}
    \caption{Mean ADHD Index by persona group and model for self-report (left) and observer-rated (right) measures. Cells show mean scores with color indicating magnitude.}
    \label{fig:model-comparison-self-and-observer}
\end{figure}

\newpage

\subsubsection{Stability by Persona and Prompt Design}

\begin{table}[H]
\centering
\begin{threeparttable}
\caption{Descriptive statistics of the between-conversation stability by persona and prompt design (Exp~I).}
\label{tab:exp1_persona_prompt}
\footnotesize
\begin{tabular}{llcccccccccc}
\toprule
& & \multicolumn{5}{c}{\textbf{Self-Report}} & \multicolumn{5}{c}{\textbf{Observer}} \\
\cmidrule(lr){3-7} \cmidrule(lr){8-12}
\textbf{Persona} & \textbf{Prompt} & $N$ & $M$ & $SD$ & Min & Max & $N$ & $M$ & $SD$ & Min & Max \\
\midrule
Default & -- & 344 & 14.7 & 4.45 & 0 & 22 & 344 & 2.19 & 1.75 & 0.00 & 9.33 \\
\midrule
\multirow{3}{*}{Low}
  & Scale & 346 & 0.85 & 1.13 & 0 & 6 & 346 & 0.16 & 0.45 & 0.00 & 3.33 \\
  & Text  & 344 & 1.78 & 1.76 & 0 & 8 & 344 & 0.90 & 3.18 & 0.00 & 18.0 \\
  & Paraphrased & 350 & 1.03 & 1.43 & 0 & 7 & 350 & 0.01 & 0.07 & 0.00 & 0.67 \\
\midrule
\multirow{3}{*}{Moderate}
  & Scale & 348 & 15.1 & 2.30 & 7 & 22 & 348 & 13.4 & 2.98 & 6.00 & 20.3 \\
  & Text  & 346 & 21.1 & 3.29 & 14 & 31 & 346 & 17.6 & 3.70 & 5.67 & 23.7 \\
  & Paraphrased & 350 & 19.3 & 3.55 & 12 & 29 & 350 & 15.9 & 3.49 & 6.33 & 23.0 \\
\midrule
\multirow{3}{*}{High}
  & Scale & 348 & 30.1 & 2.30 & 23 & 36 & 348 & 20.7 & 3.00 & 12.7 & 26.0 \\
  & Text  & 347 & 28.2 & 2.51 & 21 & 33 & 347 & 19.9 & 2.80 & 12.0 & 25.7 \\
  & Paraphrased & 350 & 28.6 & 2.55 & 24 & 36 & 350 & 19.4 & 2.12 & 13.0 & 24.0 \\
\bottomrule
\end{tabular}
\begin{tablenotes}
\footnotesize
\item \textit{Note.} Mean ($M$), Standard deviation ($SD$).
\end{tablenotes}
\end{threeparttable}
\end{table}

\newpage

\subsection{Experiment II: Additional Results}

\subsubsection{Stability by Persona and Model}

\begin{table}[H]
\centering
\begin{threeparttable}
\caption{Descriptive statistics of within-conversation stability by persona, model, and turn (Exp~II).}
\label{tab:exp2_persona_model}
\footnotesize
\begin{tabular}{lllccccccccc}
\toprule
& & & \multicolumn{4}{c}{\textbf{Self-Report}} & \multicolumn{4}{c}{\textbf{Observer}} \\
\cmidrule(lr){4-7} \cmidrule(lr){8-11}
\textbf{Persona} & \textbf{Model} & \textbf{Turn} & $N$ & $M$ & $SD$ & $\Delta$ & $N$ & $M$ & $SD$ & $\Delta$ \\
\midrule
\multirow{21}{*}{Default}
& \multirow{3}{*}{Claude}
  & 6  & 20 & 18.4 & 1.35 &  & 20 & 2.07 & 1.21 &  \\
& & 12 & 20 & 18.4 & 1.79 &  & 20 & 3.70 & 2.51 &  \\
& & 18 & 20 & 19.4 & 2.50 & +1.0 & 20 & 4.48 & 2.88 & +2.4 \\
\cmidrule(lr){2-11}
& \multirow{3}{*}{DeepSeek}
  & 6  & 20 & 10.0 & 3.09 &  & 20 & 0.53 & 0.55 &  \\
& & 12 & 20 & 10.0 & 4.73 &  & 20 & 0.42 & 0.97 &  \\
& & 18 & 20 & 8.7 & 4.57 & $-$1.3 & 20 & 0.43 & 0.62 & $-$0.1 \\
\cmidrule(lr){2-11}
& \multirow{3}{*}{Gemini}
  & 6  & 19 & 12.4 & 4.22 &  & 19 & 1.39 & 0.56 &  \\
& & 12 & 20 & 15.1 & 4.32 &  & 20 & 1.72 & 1.53 &  \\
& & 18 & 20 & 15.6 & 4.12 & +3.2 & 20 & 2.63 & 2.40 & +1.2 \\
\cmidrule(lr){2-11}
& \multirow{3}{*}{GPT 5.1}
  & 6  & 20 & 16.0 & 0.89 &  & 20 & 2.68 & 1.75 &  \\
& & 12 & 20 & 15.4 & 1.32 &  & 20 & 1.75 & 2.00 &  \\
& & 18 & 20 & 16.2 & 1.64 & +0.2 & 20 & 1.65 & 1.62 & $-$1.0 \\
\cmidrule(lr){2-11}
& \multirow{3}{*}{GPT OSS}
  & 6  & 14 & 15.7 & 4.63 &  & 20 & 0.87 & 0.46 &  \\
& & 12 & 16 & 16.7 & 1.62 &  & 19 & 1.07 & 0.61 &  \\
& & 18 & 20 & 17.2 & 1.15 & +1.5 & 20 & 1.22 & 0.51 & +0.4 \\
\cmidrule(lr){2-11}
& \multirow{3}{*}{Grok}
  & 6  & 8 & 16.6 & 2.20 &  & 14 & 1.60 & 1.04 &  \\
& & 12 & 6 & 18.5 & 1.87 &  & 14 & 1.45 & 1.34 &  \\
& & 18 & 14 & 16.5 & 4.09 & $-$0.1 & 14 & 1.19 & 0.83 & $-$0.4 \\
\cmidrule(lr){2-11}
& \multirow{3}{*}{Llama}
  & 6  & 20 & 16.3 & 1.63 &  & 20 & 0.62 & 0.45 &  \\
& & 12 & 20 & 15.1 & 2.47 &  & 20 & 0.52 & 0.64 &  \\
& & 18 & 20 & 13.8 & 3.12 & $-$2.5 & 20 & 0.73 & 0.82 & +0.1 \\
\midrule
\multirow{21}{*}{Low}
& \multirow{3}{*}{Claude}
  & 6  & 58 & 1.19 & 0.98 &  & 58 & 0.09 & 0.23 &  \\
& & 12 & 59 & 1.61 & 1.17 &  & 59 & 0.25 & 0.42 &  \\
& & 18 & 60 & 1.45 & 1.14 & +0.3 & 60 & 0.47 & 0.54 & +0.4 \\
\cmidrule(lr){2-11}
& \multirow{3}{*}{DeepSeek}
  & 6  & 59 & 0.59 & 0.85 &  & 59 & 0.05 & 0.17 &  \\
& & 12 & 59 & 0.34 & 0.78 &  & 59 & 0.06 & 0.17 &  \\
& & 18 & 59 & 0.31 & 0.73 & $-$0.3 & 59 & 0.08 & 0.28 & 0.0 \\
\cmidrule(lr){2-11}
& \multirow{3}{*}{Gemini}
  & 6  & 58 & 0.02 & 0.13 &  & 58 & 0.10 & 0.17 &  \\
& & 12 & 57 & 0.02 & 0.13 &  & 57 & 0.11 & 0.19 &  \\
& & 18 & 58 & 0.07 & 0.26 & +0.1 & 58 & 0.13 & 0.18 & 0.0 \\
\cmidrule(lr){2-11}
& \multirow{3}{*}{GPT 5.1}
  & 6  & 60 & 0.97 & 1.34 &  & 59 & 0.12 & 0.28 &  \\
& & 12 & 60 & 1.23 & 1.37 &  & 60 & 0.16 & 0.34 &  \\
& & 18 & 60 & 1.12 & 1.43 & +0.2 & 60 & 0.22 & 0.64 & +0.1 \\
\cmidrule(lr){2-11}
& \multirow{3}{*}{GPT OSS}
  & 6  & 53 & 0.60 & 1.34 &  & 53 & 0.25 & 0.23 &  \\
& & 12 & 51 & 0.63 & 1.13 &  & 54 & 0.51 & 0.31 &  \\
& & 18 & 55 & 1.02 & 1.83 & +0.4 & 55 & 0.78 & 0.59 & +0.5 \\
\cmidrule(lr){2-11}
& \multirow{3}{*}{Grok}
  & 6  & 60 & 0.58 & 1.08 &  & 60 & 0.03 & 0.09 &  \\
& & 12 & 60 & 0.70 & 0.98 &  & 60 & 0.08 & 0.17 &  \\
& & 18 & 60 & 0.83 & 1.26 & +0.3 & 60 & 0.06 & 0.15 & 0.0 \\
\cmidrule(lr){2-11}
& \multirow{3}{*}{Llama}
  & 6  & 59 & 1.34 & 1.55 &  & 58 & 0.05 & 0.13 &  \\
& & 12 & 59 & 1.69 & 1.50 &  & 58 & 0.10 & 0.26 &  \\
& & 18 & 59 & 1.76 & 1.49 & +0.4 & 58 & 0.31 & 0.40 & +0.3 \\
\bottomrule
\end{tabular}
\begin{tablenotes}
\footnotesize
\item \textit{Note.} Table continues on next page.
\end{tablenotes}
\end{threeparttable}
\end{table}

\begin{table}[H]
\centering
\begin{threeparttable}
\caption{Descriptive statistics of within-conversation stability by persona, model, and turn (Exp~II, continued).}
\label{tab:exp2_persona_model_cont}
\footnotesize
\begin{tabular}{lllccccccccc}
\toprule
& & & \multicolumn{4}{c}{\textbf{Self-Report}} & \multicolumn{4}{c}{\textbf{Observer}} \\
\cmidrule(lr){4-7} \cmidrule(lr){8-11}
\textbf{Persona} & \textbf{Model} & \textbf{Turn} & $N$ & $M$ & $SD$ & $\Delta$ & $N$ & $M$ & $SD$ & $\Delta$ \\
\midrule
\multirow{21}{*}{Moderate}
& \multirow{3}{*}{Claude}
  & 6  & 60 & 16.3 & 2.79 &  & 59 & 14.5 & 5.79 &  \\
& & 12 & 60 & 16.3 & 3.14 &  & 60 & 14.2 & 4.82 &  \\
& & 18 & 60 & 16.3 & 3.13 & 0.0 & 58 & 13.8 & 4.72 & $-$0.7 \\
\cmidrule(lr){2-11}
& \multirow{3}{*}{DeepSeek}
  & 6  & 58 & 14.8 & 4.18 &  & 58 & 10.3 & 3.11 &  \\
& & 12 & 56 & 14.9 & 4.59 &  & 56 & 8.98 & 3.50 &  \\
& & 18 & 58 & 14.7 & 3.94 & $-$0.1 & 57 & 8.00 & 3.13 & $-$2.3 \\
\cmidrule(lr){2-11}
& \multirow{3}{*}{Gemini}
  & 6  & 58 & 17.8 & 7.50 &  & 57 & 14.8 & 4.20 &  \\
& & 12 & 58 & 20.2 & 7.75 &  & 58 & 15.7 & 3.86 &  \\
& & 18 & 59 & 19.4 & 8.15 & +1.6 & 59 & 15.1 & 3.81 & +0.3 \\
\cmidrule(lr){2-11}
& \multirow{3}{*}{GPT 5.1}
  & 6  & 60 & 15.6 & 2.66 &  & 60 & 13.4 & 2.50 &  \\
& & 12 & 60 & 14.8 & 2.88 &  & 58 & 9.78 & 3.49 &  \\
& & 18 & 60 & 14.9 & 2.53 & $-$0.7 & 59 & 8.69 & 3.94 & $-$4.7 \\
\cmidrule(lr){2-11}
& \multirow{3}{*}{GPT OSS}
  & 6  & 58 & 19.3 & 4.93 &  & 59 & 14.0 & 2.74 &  \\
& & 12 & 59 & 18.7 & 4.15 &  & 58 & 10.9 & 2.29 &  \\
& & 18 & 60 & 18.8 & 4.94 & $-$0.5 & 60 & 10.8 & 1.93 & $-$3.2 \\
\cmidrule(lr){2-11}
& \multirow{3}{*}{Grok}
  & 6  & 56 & 17.6 & 4.15 &  & 58 & 12.7 & 3.61 &  \\
& & 12 & 57 & 18.2 & 4.26 &  & 58 & 10.9 & 3.60 &  \\
& & 18 & 58 & 17.0 & 3.50 & $-$0.6 & 58 & 9.84 & 4.00 & $-$2.9 \\
\cmidrule(lr){2-11}
& \multirow{3}{*}{Llama}
  & 6  & 60 & 16.6 & 0.83 &  & 60 & 12.8 & 2.95 &  \\
& & 12 & 59 & 16.4 & 0.55 &  & 58 & 11.1 & 2.68 &  \\
& & 18 & 60 & 16.4 & 0.52 & $-$0.2 & 60 & 9.19 & 2.78 & $-$3.6 \\
\midrule
\multirow{21}{*}{High}
& \multirow{3}{*}{Claude}
  & 6  & 60 & 27.6 & 1.72 &  & 60 & 20.3 & 1.44 &  \\
& & 12 & 59 & 28.5 & 1.78 &  & 57 & 19.9 & 1.96 &  \\
& & 18 & 60 & 28.6 & 1.54 & +1.0 & 60 & 18.7 & 2.07 & $-$1.6 \\
\cmidrule(lr){2-11}
& \multirow{3}{*}{DeepSeek}
  & 6  & 59 & 31.6 & 3.31 &  & 58 & 15.0 & 2.07 &  \\
& & 12 & 59 & 31.4 & 3.08 &  & 58 & 13.9 & 2.80 &  \\
& & 18 & 59 & 31.2 & 3.70 & $-$0.4 & 59 & 11.9 & 2.81 & $-$3.1 \\
\cmidrule(lr){2-11}
& \multirow{3}{*}{Gemini}
  & 6  & 55 & 29.4 & 1.25 &  & 56 & 19.7 & 1.80 &  \\
& & 12 & 55 & 29.6 & 1.63 &  & 56 & 19.4 & 2.01 &  \\
& & 18 & 58 & 29.8 & 1.67 & +0.4 & 57 & 18.4 & 2.80 & $-$1.3 \\
\cmidrule(lr){2-11}
& \multirow{3}{*}{GPT 5.1}
  & 6  & 58 & 25.7 & 1.99 &  & 58 & 17.5 & 1.97 &  \\
& & 12 & 59 & 26.2 & 2.15 &  & 59 & 13.5 & 2.38 &  \\
& & 18 & 59 & 25.9 & 1.96 & +0.2 & 59 & 12.0 & 2.73 & $-$5.5 \\
\cmidrule(lr){2-11}
& \multirow{3}{*}{GPT OSS}
  & 6  & 58 & 27.4 & 2.42 &  & 57 & 17.1 & 1.35 &  \\
& & 12 & 58 & 27.7 & 2.65 &  & 58 & 13.7 & 2.25 &  \\
& & 18 & 58 & 27.9 & 2.63 & +0.5 & 58 & 12.7 & 2.02 & $-$4.4 \\
\cmidrule(lr){2-11}
& \multirow{3}{*}{Grok}
  & 6  & 55 & 30.3 & 1.46 &  & 56 & 15.9 & 1.47 &  \\
& & 12 & 56 & 29.8 & 1.37 &  & 56 & 12.9 & 2.65 &  \\
& & 18 & 56 & 29.8 & 1.53 & $-$0.5 & 56 & 12.4 & 2.85 & $-$3.5 \\
\cmidrule(lr){2-11}
& \multirow{3}{*}{Llama}
  & 6  & 60 & 25.6 & 1.34 &  & 60 & 16.6 & 1.92 &  \\
& & 12 & 59 & 25.3 & 1.73 &  & 59 & 14.3 & 1.95 &  \\
& & 18 & 60 & 25.5 & 1.64 & $-$0.1 & 60 & 11.8 & 2.47 & $-$4.8 \\
\bottomrule
\end{tabular}
\begin{tablenotes}
\footnotesize
\item \textit{Note.} Mean ($M$), Standard deviation ($SD$), $\Delta$ = Mean change from turn 6 to 18.
\end{tablenotes}
\end{threeparttable}
\end{table}

\newpage

\subsubsection{Stability by Persona and Prompt Design}

\begin{table}[H]
\centering
\begin{threeparttable}
\caption{Descriptive statistics of within-conversation stability by persona, prompt design, and turn (Exp~II).}
\label{tab:exp2_persona_prompt}
\footnotesize
\begin{tabular}{lllccccccccccc}
\toprule
& & & \multicolumn{5}{c}{\textbf{Self-Report}} & & \multicolumn{5}{c}{\textbf{Observer}} \\
\cmidrule(lr){4-8} \cmidrule(lr){10-14}
\textbf{Persona} & \textbf{Prompt} & \textbf{Turn} & $N$ & $M$ & $SD$ & $\Delta$ & 
& & $N$ & $M$ & $SD$ & $\Delta$  \\
\midrule
\multirow{3}{*}{Default}
& \multirow{3}{*}{--}
  & 6  & 121 & 14.9 & 3.92 & & & & 133 & 1.38 & 1.21 & & \\
& & 12 & 122 & 15.2 & 3.94 & & & & 133 & 1.52 & 1.83 & & \\
& & 18 & 134 & 15.3 & 4.48 & +0.4 & & & 134 & 1.79 & 2.09 & +0.4 &  \\
\midrule
\multirow{9}{*}{Low}
& \multirow{3}{*}{Scale}
  & 6  & 137 & 0.51 & 1.04 & & & & 137 & 0.11 & 0.22 & & \\
& & 12 & 135 & 0.56 & 0.96 & & & & 137 & 0.15 & 0.29 & & \\
& & 18 & 137 & 0.72 & 1.40 & +0.2 & & & 137 & 0.19 & 0.42 & +0.1 &  \\
\cmidrule(lr){2-14}
& \multirow{3}{*}{Text}
  & 6  & 138 & 1.42 & 1.40 & & & & 137 & 0.09 & 0.19 & & \\
& & 12 & 135 & 1.56 & 1.44 & & & & 136 & 0.18 & 0.31 & & \\
& & 18 & 138 & 1.53 & 1.46 & +0.1 & & & 138 & 0.37 & 0.57 & +0.3 &  \\
\cmidrule(lr){2-14}
& \multirow{3}{*}{Paraphrased}
  & 6  & 132 & 0.33 & 0.70 & & & & 131 & 0.08 & 0.20 & & \\
& & 12 & 135 & 0.58 & 0.98 & & & & 134 & 0.20 & 0.34 & & \\
& & 18 & 136 & 0.57 & 0.99 & +0.2 & & & 135 & 0.30 & 0.47 & +0.2 &  \\
\midrule
\multirow{9}{*}{Moderate}
& \multirow{3}{*}{Scale}
  & 6  & 139 & 12.5 & 3.20 & & & & 139 & 9.10 & 2.35 & & \\
& & 12 & 137 & 12.6 & 3.02 & & & & 136 & 8.41 & 2.91 & & \\
& & 18 & 140 & 12.5 & 3.13 & 0.0 & & & 138 & 7.81 & 3.29 & $-$1.3 &  \\
\cmidrule(lr){2-14}
& \multirow{3}{*}{Text}
  & 6  & 134 & 18.6 & 2.84 & & & & 135 & 14.8 & 2.97 & & \\
& & 12 & 136 & 18.9 & 3.75 & & & & 135 & 12.8 & 3.89 & & \\
& & 18 & 137 & 18.7 & 3.51 & +0.1 & & & 135 & 12.1 & 3.86 & $-$2.7 &  \\
\cmidrule(lr){2-14}
& \multirow{3}{*}{Paraphrased}
  & 6  & 137 & 19.5 & 3.54 & & & & 137 & 15.8 & 2.42 & & \\
& & 12 & 136 & 19.7 & 3.67 & & & & 135 & 13.9 & 3.47 & & \\
& & 18 & 138 & 19.2 & 3.88 & $-$0.3 & & & 138 & 12.5 & 4.14 & $-$3.3 &  \\
\midrule
\multirow{9}{*}{High}
& \multirow{3}{*}{Scale}
  & 6  & 134 & 29.9 & 2.38 & & & & 133 & 17.3 & 2.41 & & \\
& & 12 & 133 & 29.8 & 2.41 & & & & 134 & 15.7 & 3.67 & & \\
& & 18 & 136 & 29.6 & 2.62 & $-$0.3 & & & 136 & 14.6 & 4.09 & $-$2.7 &  \\
\cmidrule(lr){2-14}
& \multirow{3}{*}{Text}
  & 6  & 138 & 26.9 & 2.76 & & & & 137 & 17.3 & 2.47 & & \\
& & 12 & 138 & 27.2 & 2.89 & & & & 137 & 14.9 & 3.54 & & \\
& & 18 & 139 & 27.5 & 2.95 & +0.6 & & & 139 & 13.3 & 3.65 & $-$4.0 &  \\
\cmidrule(lr){2-14}
& \multirow{3}{*}{Paraphrased}
  & 6  & 133 & 27.8 & 2.79 & & & & 135 & 17.8 & 2.57 & & \\
& & 12 & 134 & 28.1 & 2.76 & & & & 132 & 15.4 & 3.46 & & \\
& & 18 & 135 & 28.1 & 2.85 & +0.3 & & & 134 & 14.1 & 3.76 & $-$3.7 &  \\
\bottomrule
\end{tabular}
\begin{tablenotes}
\footnotesize
\item \textit{Note.} Mean ($M$), Standard deviation ($SD$), $\Delta$ = Mean change from turn 6 to 18.  
\end{tablenotes}
\end{threeparttable}
\end{table}

\newpage

\subsection{Inter-Rater Reliability Details}

\begin{table}[h]
\centering
\caption{Mean observer ratings by evaluator LLM and persona group.}
\label{tab:app_evaluator_means}
\footnotesize
\begin{tabular}{llcccc}
\toprule
& & \multicolumn{2}{c}{\textbf{Exp~I}} & \multicolumn{2}{c}{\textbf{Exp~II}} \\
\cmidrule(lr){3-4} \cmidrule(lr){5-6}
\textbf{Evaluator} & \textbf{Persona} & $M$ & $SD$ & $M$ & $SD$ \\
\midrule
\multirow{4}{*}{Claude Sonnet 4.5}
  & Default      & 5.17 & 3.67 & 3.54 & 2.91 \\
  & High         & 22.3 & 2.40 & 19.0 & 3.60 \\
  & Moderate     & 17.9 & 4.00 & 15.0 & 4.58 \\
  & Low & 0.51 & 2.26 & 0.51 & 0.92 \\
\midrule
\multirow{4}{*}{Gemini 3 Pro}
  & Default      & 0.54 & 1.51 & 0.89 & 2.22 \\
  & High         & 21.4 & 3.24 & 17.8 & 4.08 \\
  & Moderate     & 16.9 & 4.17 & 13.9 & 5.06 \\
  & Low & 0.30 & 2.06 & 0.03 & 0.31 \\
\midrule
\multirow{4}{*}{GPT 5.1}
  & Default      & 0.86 & 1.27 & 0.26 & 0.79 \\
  & High         & 16.2 & 3.19 & 10.0 & 4.27 \\
  & Moderate     & 12.0 & 3.89 & 6.80 & 4.08 \\
  & Low & 0.26 & 1.42 & 0.02 & 0.18 \\
\bottomrule
\end{tabular}
\end{table}

\end{document}